\documentclass[]{spie}  

\usepackage{amsmath,amsfonts,amssymb}
\usepackage{graphicx}
\usepackage[colorlinks=true, allcolors=blue]{hyperref}

\newcommand\norm[1]{\left\lVert#1\right\rVert}

\usepackage{url}

\title{Unsupervised Domain Adaptation via CycleGAN for White Matter Hyperintensity Segmentation in Multicenter MR Images}

\author[a]{Julian Alberto Palladino}
\author[a]{Diego Fernandez Slezak}
\author[b]{Enzo Ferrante}
\affil[a]{Laboratorio de Inteligencia Artificial Aplicada, Instituto de Ciencias de la Computación, CONICET, Universidad de Buenos Aires, Buenos Aires, Argentina}
\affil[b]{Research institute for
signals, systems and
computational intelligence, sinc(i),
CONICET, Universidad Nacional del Litoral, Santa Fe, Argentina}

\authorinfo{Further author information:\\
J.A.P.: E-mail: julianpalladino8@gmail.com\\
D.F.S.: E-mail: dfslezak@dc.uba.ar\\
E.F.: E-mail: eferrante@sinc.unl.edu.ar}

\begin{document} 
\maketitle

\begin{abstract}
Automatic segmentation of white matter hyperintensities in magnetic resonance images is of paramount clinical and research importance. Quantification of these lesions serve as a predictor for risk of stroke, dementia and mortality. 
During the last years, convolutional neural networks (CNN) specifically tailored for biomedical image segmentation have outperformed all previous techniques in this task. However, they are extremely data-dependent, and maintain a good performance only when data distribution between training and test datasets remains unchanged. When such distribution changes but we still aim at performing the same task, we incur in a domain adaptation problem (e.g. using a different MR machine or different acquisition parameters for training and test data). In this work, we explore the use of cycle-consistent adversarial networks (CycleGAN) to perform unsupervised domain adaptation on multicenter MR images with brain lesions. We aim at learning a mapping function to transform volumetric MR images between domains, which are characterized by different medical centers and MR machines with varying brand, model and configuration parameters. Our experiments show that CycleGAN allows us to reduce the Jensen-Shannon divergence between MR domains, enabling automatic segmentation with CNN models on domains where no labeled data was available. 
\end{abstract}

\section{Introduction}
 White matter hyperintensities (WMH), also known as leukoaraiosis, are a characteristic of small vessel disease commonly observed in the brain of elderly subjects \cite{Guerrero2017}. Magnetic resonance image (MRI) is the modality of choice to study WMH lesions. During the last years, it has been shown that accurate quantification of WMH volume is of paramount clinical importance since it may serve as a predictor for risk of stroke, dementia and mortality \cite{Debette2010}. 
 Given that manual delineation of these lesions is a difficult and time consuming task, several computational methods were recently proposed to deal with automatic WHM segmentation. CNN architectures specifically tailored for biomedical image segmentation \cite{unet2d,kamnitsas2017efficient,shakeri2016sub} have outperformed all previous techniques in the task of automatic brain structures segmentation in general, and WMH in particular \cite{Guerrero2017,Rachmadi2018,Kuijf2019,Roulet2019}.
 However, these models are extremely data-dependent, in the sense that they require many annotated images to be trained. More importantly, they maintain a good performance only when the data distribution between training (source) and test (target) domains remains unchanged. When such distribution changes (incurring in a co-variate shift scenario \cite{Storkey2009}) but we still aim at performing the same task, domain adaptation techniques \cite{Pan2009} can be used to achieve better performance in unseen target domains.
 
 In the context of WMH segmentation in MRI,  co-variate shift and domain adaptation problems may arise when we have trained a model with images coming from a particular medical center, MR machine brand or parameter setup, and we want to test it on images acquired under different conditions. In this case, the performance of the segmentation algorithm tends to decrease. Several studies have empirically shown this behaviour and proposed alternative methods to deal with it. Ghafoorian and co-workers \cite{Ghafoorian2017} showed that it is possible to apply supervised transfer learning to re-use WMH segmentation models when annotated images are available in the target domain. In this case, simple fine-tunning of a previously trained model is enough to achieve state-of-the-art results in the new domain. The disadvantage of this supervised approach is that we require manual annotations in the target domain.

In this work we focus on strategies which do not require manual annotations for the target domain. Following this idea, several approaches based on adversarial training have been proposed. The work of Kamnitsas et al \cite{Kamnitsas2017} was one of the first ones employing adversarial training to learn domain invariant features for the task of brain lesion segmentation. More recently, Orbes-Orteaga \cite{Orbes2019} proposed a different strategy which also employs adversarial learning but they combine it with a consistency loss term requiring multiple target domains with paired images, which is not a common situation in clinical scenarios. Moreover, both adversarial approaches require to have access to the unlabeled images during training, making it difficult to apply the resulting models in completely unseen domains. Here we will focus on learning a mapping function to shift the target distribution towards the source distribution, so that previously trained models can be directly applied in new scenarios.

Closest to our work are those of \cite{romobucheli2019} and \cite{Liu2019}, which pose domain adaptation as an image translation problem and employ Cycle-Consistent Adversarial Networks (CycleGAN) \cite{Zhu2017} to translate from target to source domain. Differently from us, \cite{romobucheli2019} focuses on optical coherence tomography (OCT) images while \cite{Liu2019} explores the use of CycleGAN for anatomical segmentation (bilateral amygdala) in brain MR. \\

\noindent \textbf{Contributions:} To the best of our knowleadge, our work is the first one to provide empirical evidence that CycleGAN enables segmentation in multicentric MR data for brain lesions. To this end, we study one of the most challenging brain lesion segmentation problems, namely WMH. In addition, while previous approaches employ cycle-consistent adversarial networks operating only on 2D patches, we show that it is possible to train them directly operating on tridimensional images. We measure the effectiveness of our approach by analyzing not only the multicenter segmentation results, but also the co-variate shift in terms of pairwise Jensen-Shannon divergences after domain mapping. We show that lower inter-domain Jensen-Shannon divergences correlate with better performance across different domains. Our experimental evaluation in brain MR images coming from three different medical centers demonstrate that unsupervised domain adaptation via CycleGAN improves WMH segmentation in multicenter MR images.

\begin{figure}[t!]
 \centering
 \includegraphics[width=0.95\textwidth]{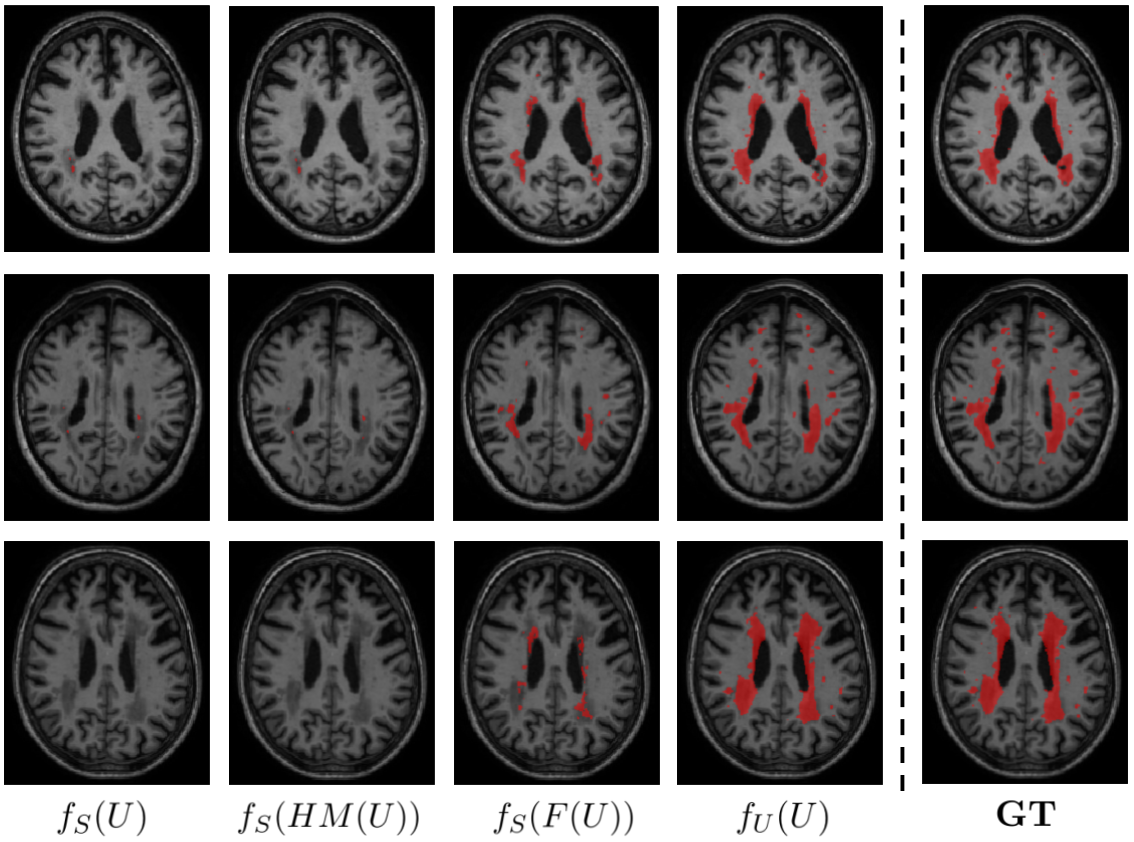}
 \caption{Qualitative results for WMH segmentation (in red). Examples considering Singapore ($S$) as source and Utrecht ($U$) as target domain. $f_S$ and $f_U$ are the segmentation models trained on Singapore and Utrecht respectively. From left to right: (i) no domain adaptation; (ii) adaptation via histogram matching; (iii) adaptation via CycleGAN (iv) training in target domain and (v) ground-truth.}
 \label{fig:visualResults}
\end{figure}

\section{Unsupervised Domain Adaptation via CycleGAN}
We highlight that the main contribution of this work is not related to a novel generative adversarial network. Instead, we aim at providing empirical evidence that existing Cycle-GANs tailored to process 3D images help to perform domain adaptation in the context of brain lesion segmentation for multicenter MR data. For completeness, we include a brief description of the Cycle-GAN framework.

\begin{figure}
 \centering
    \includegraphics[width=0.7\textwidth]{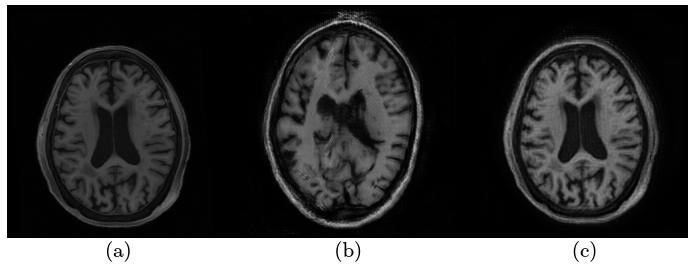}
 \caption{Effect of the identity mapping regularization term when training CycleGAN. (a) Input image. (b) Hallucinations induced by a CycleGAN trained without identity mapping. (c) Results obtained with CycleGAN trained with identity mapping (note there is no change in morphology, only in the intesities).}
 \label{fig:id_mapping}
\end{figure}

Cycle-GAN is a style-transfer CNN model based on generative adversarial networks \cite{goodfellow2014generative}, but redesigned with the specific goal of translating images from a source domain $X$ to a target domain $Y$ in the absence of paired examples. The idea is to learn a mapping function $G:X \rightarrow Y$ such that the distribution of images from $G(X)$ is indistinguishable from the distribution $Y$ by using an adversarial loss. Because this mapping is highly under-constrained, it is coupled with an inverse mapping $F: Y \rightarrow X$, thus introducing a cycle consistency loss to enforce $F(G(X)) \approx X$ (and vice versa). The functions $G$ and $F$ are neural networks which follow a encoder-decoder architecture (see Appendix section \ref{app:generator} for a detailed description of the generator architecture). The framework also incorporates discriminators $D_X$ and $D_Y$ which learn to distinguish between translated and real examples following a standard adversarial scheme \cite{goodfellow2014generative} (see Appendix section \ref{app:discriminator} for a detailed description of the discriminator architecture) . Additionally, an identity mapping term is introduced in the loss function to encourage $F$ and $G$ to apply the identity transformation when real samples of the target domain are provided as the input to the generator. The identity regularization plays a crucial role in producing realistic mapping functions and avoiding potential hallucinations that may emerge during image translation (see Figure \ref{fig:id_mapping} for a visual example of such hallucinations). In the following we describe the terms included in the final loss function used to train the CycleGAN.\\

\noindent \textbf{Adversarial loss.}
The adversarial term encourages the mapping functions to translate from one domain to the other. It is applied to both mapping functions $F$ and $G$. For $G:X \rightarrow Y$ and its corresponding adversarial discriminator $D_Y$, the objective can be expressed as:
\begin{equation*}
    \mathcal{L}_{GAN}(G,D_{Y},X,Y) = \mathbb{E}_{y \sim p_{data}(y)} [log( D_{Y}(y))]  
      + \mathbb{E}_{x \sim p_{data}(x)} [log (1-D_{Y}(G(x)))],
\end{equation*}
\noindent where G generates images $G(x)$ that look similar to images from domain $Y$, while $D_Y$ learns to distinguish between translated samples $G(x)$ and real samples $y$. $G$ aims to minimize this objective against its adversary $D_Y$, that tries to maximize it, i.e., min$_G$ max$_{D_Y} \mathcal{L}_{GAN} (G,D_Y,X,Y)$. An analogous loss is also introduced for $F:Y \rightarrow X$ and its discriminator $D_X$ as follows:
\begin{equation*}
    \mathcal{L}_{GAN}(F,D_{X},Y,X) = \mathbb{E}_{x \sim p_{data}(x)} [log(D_{S}(x))] + \mathbb{E}_{y \sim p_{data}(y)} [log (1-D_{X}(F(y)))],
\end{equation*}

\noindent where $F(y)$ and $D_X$ behave analogously, i.e., min$_F$ max$_{D_X} \mathcal{L}_{GAN} (F,D_X,Y,X)$.\\

\noindent \textbf{Cycle consistency loss.} Adversarial losses alone do not guarantee that images can be converted back and forth from X to Y and vice versa. Intuitively, if an image $x_i \in X$ is mapped to a domain $Y$ by $G(x_i)$, applying the inverse mapping $F(G(x_i))$ should return the exact same image $x_i$. This behaviour is encouraged by the cycle consistency term and is formulated as:
\begin{equation*}
    \mathcal{L}_{cyc}(G,F) =
    \mathbb{E}_{x \sim p_{data}(x)} [  \norm{F(G(x))-x}_{1}] 
     + \mathbb{E}_{y \sim p_{data}(y)} [\norm{G(F(y))-y}_{1}]
\end{equation*}

\noindent \textbf{Identity mapping loss.}
Last but not least, the identity mapping term encourages $F$ and $G$ to apply the identity transformation when real samples of the target domain are provided as the input to the generator. This behaviour is encoded in the following equation:
\begin{equation*}
    \mathcal{L}_{id}(G,F) = \mathbb{E}_{x \sim p_{data}(x)} [\norm{F(x)-x}_{1}] 
    + \mathbb{E}_{y \sim p_{data}(y)} [\norm{G(y)-y}_{1}]
\end{equation*}

\begin{figure}[t!]
 \centering
    \includegraphics[width=0.6\textwidth]{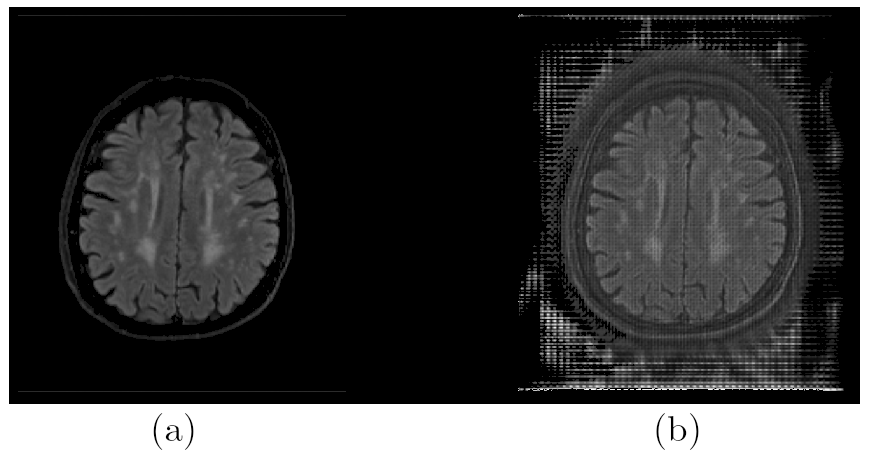}
 \caption{Example of the ``checkerboard artifact''. (a) correctly transformed image (obtained with a generator which uses resize convolution). (b) ``checkerboard artifact' obtained when using transposed convolutions.}
 \label{fig:resize_conv}
\end{figure}

\noindent \textbf{CycleGAN Training.} The final loss function used to train the CycleGAN model is defined as the sum of the adversarial (${L}_{GAN}$), cycle consistency ($\mathcal{L}_{cyc}$) and identity (${L}_{id}$) losses. 
The model is trained following an iterative approach, where each step consists in the training of the discriminators over one real image and one synthetic image, followed by the generators trained to translate one instance each. We define an epoch of the whole training process as 1000 of these steps, and the whole training lasts for 200 epochs. We adopt Adam optimizer with standard parameters and initial learning rate of 0.0002. 
We used the original CycleGAN architecture which was only modified in two ways: first, we adapted it to process 3D patches by using standard 3D convolutions. Second, we replaced transposed convolutions by resize convolutions \cite{Odena2016} in order to avoid checkerboard artifacts in the output images (see Figure \ref{fig:resize_conv}).\\

\noindent \textbf{Segmentation model.} For WMH segmentation, we employ a 3D U-Net architecture with a final softmax layer producing a lesion probability map (see Appendix section \ref{app:unet} for a detailed description of the U-Net architecture used in this work). For optimization, we used the Adam optimizer with a learning rate of 0.0002. Patch-based training is performed by constructing balanced mini-batches of image patches of $(32\times32\times32)$. We balance the mini-batches by sampling with equal probability from those patches centered on a voxel with WMH presence and those centered on a healthy voxel.\\

\noindent \textbf{Domain Adaptation.} For a source domain $X$ with ground-truth annotations, we train a segmentation model $f_X$. Then, given an unseen domain $Y$, we learn a mapping function $F: Y \rightarrow X$ using the CycleGAN framework. In this way, we enable segmentation of images from the target domain $y_i \in Y$, transforming them before segmentation. The final segmentation maps are obtained by $f_X(F(y_i))$. 

\section{Experiments}
\noindent \textbf{Database.} We employ the 2017 WMH Segmentation Challenge dataset \cite{Kuijf2019} which is publicly available and includes multicenter images. This database provides 60 brain magnetic resonance images (T1 and FLAIR sequences) captured in three different medical centers alongside their manual WMH segmentations. The 60 MRIs are divided in 3 groups: 
\begin{itemize}
    \item  \textit{University Medical Center, Utrecht}: 20 MR images captured with a 3T Philips Achieva machine. It includes T1 (Voxel size: $1.00\times1.00\times1.00 mm^{3}$. TR/TE: 7.9/4.5 ms.) and FLAIR (Voxel size: $0.96\times0.95\times3.00 mm^{3}$. TR/TE/TI: 11000/125/2800 ms.) images.
    \item \textit{National University Health System, Singapore:} 20 MR images captured with a 3T Siemens TrioTim machine. It includes T1 (Voxel size: $1.00\times1.00\times1.00 mm^{3}$. TR/TE/TI: 2300/1.9/900 ms) and FLAIR (Voxel size: $1.0\times1.0\times3.00 mm^{3}$. TR/TE/TI: 9000/82/2500 ms) images.
    \item \textit{Vrije Universiteit, Amsterdam:} 20 MR images captured with a 3T GE Signa HDxt machine. It includes T1 (Voxel size: $0.87\times0.87\times1.00 mm^3$. TR/TE: 9.9/4.6 ms) and FLAIR (Voxel size: $1.04\times1.04\times0.56 mm^3$. TR/TE/TI: 4800/279/1650 ms) images.
\end{itemize}

\noindent \textbf{Baseline histogram matching (HM) method.} For comparison, we implemented a baseline adaptation method using standard histogram matching \cite{nyul2000new}. We adopted a pairwise strategy proceeding as follows: given an image $y_i \in Y$ from the unseen target domain $Y$, we look for the image $x_i \in X$ in the training source domain $X$ that is most similar to $y_i$ in terms of Jensen-Shannon (JS) divergence \cite{lin1991divergence} (see next paragraph). We then transform the histogram of $y_i$ to match that of $x_i$ using the SimpleITK \cite{lowekamp2013design} histogram matching function.\\

\noindent \textbf{Jensen-Shannon Divergence.} JS divergence is a symmetric measure that quantifies how different are two probability distributions (see \cite{lin1991divergence} for more details about the definition of JS divergence). We interpret the histogram of intensities of every image $hist(x_i)$ as a distribution, and use the JS divergence to measure pairwise distances. Note that $hist(x_i)$ only considers the voxel intensities withing the head mask (i.e. excluding background). In this study, we employ pairwise JS divergences for two different tasks. On the one hand, as described in the previous paragraph, we use it to choose the closest image for histogram matching.

On the other hand, we use it as an indicator to quantify co-variate shift between domains. To this end, we define the average inter-domain JS divergence between all possible pairs of images from two domains $X$ and $Y$ as:
\begin{equation*}
\Delta_{JS}(X,Y) = \frac{1}{|X|.|Y|} \sum_{i=1}^{|X|} \sum_{j=1}^{|Y|} JS(hist(x_{i}),hist(y_{j})),
\end{equation*}

\noindent where $|X|$ and $|Y|$ indicate the number of images in each domain. We also define the average JS divergence intra-domain $\Delta_{JS}(X,X)$, but of course we exclude comparisons of a given image with itself ($x_i$, $x_i$). We employ inter and intra-domain pairwise JS divergences as an indicator of co-variate domain shift.
\\

\begin{figure}[t!]
 \centering
    \includegraphics[width=\textwidth]{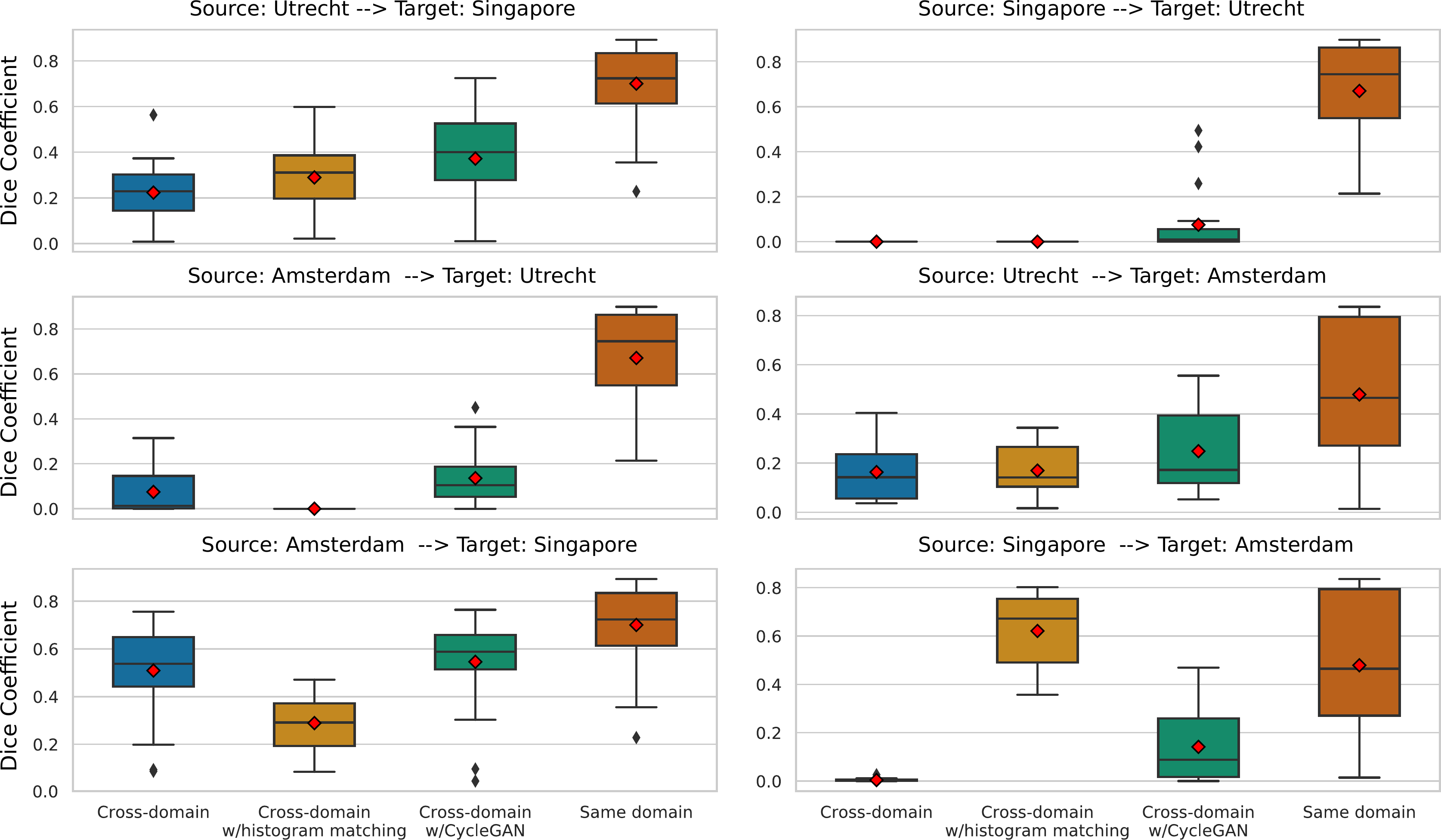}
 \caption{WHM segmentation results measured in terms of Dice for different source and target domains: (i) no domain adaptation; (ii) adaptation via histogram matching; (iii) adaptation via CycleGAN (iv) training in the target domain.}
 \label{fig:dice}
\end{figure}

\begin{figure}[t!]
 \centering
 \includegraphics[width=\textwidth]{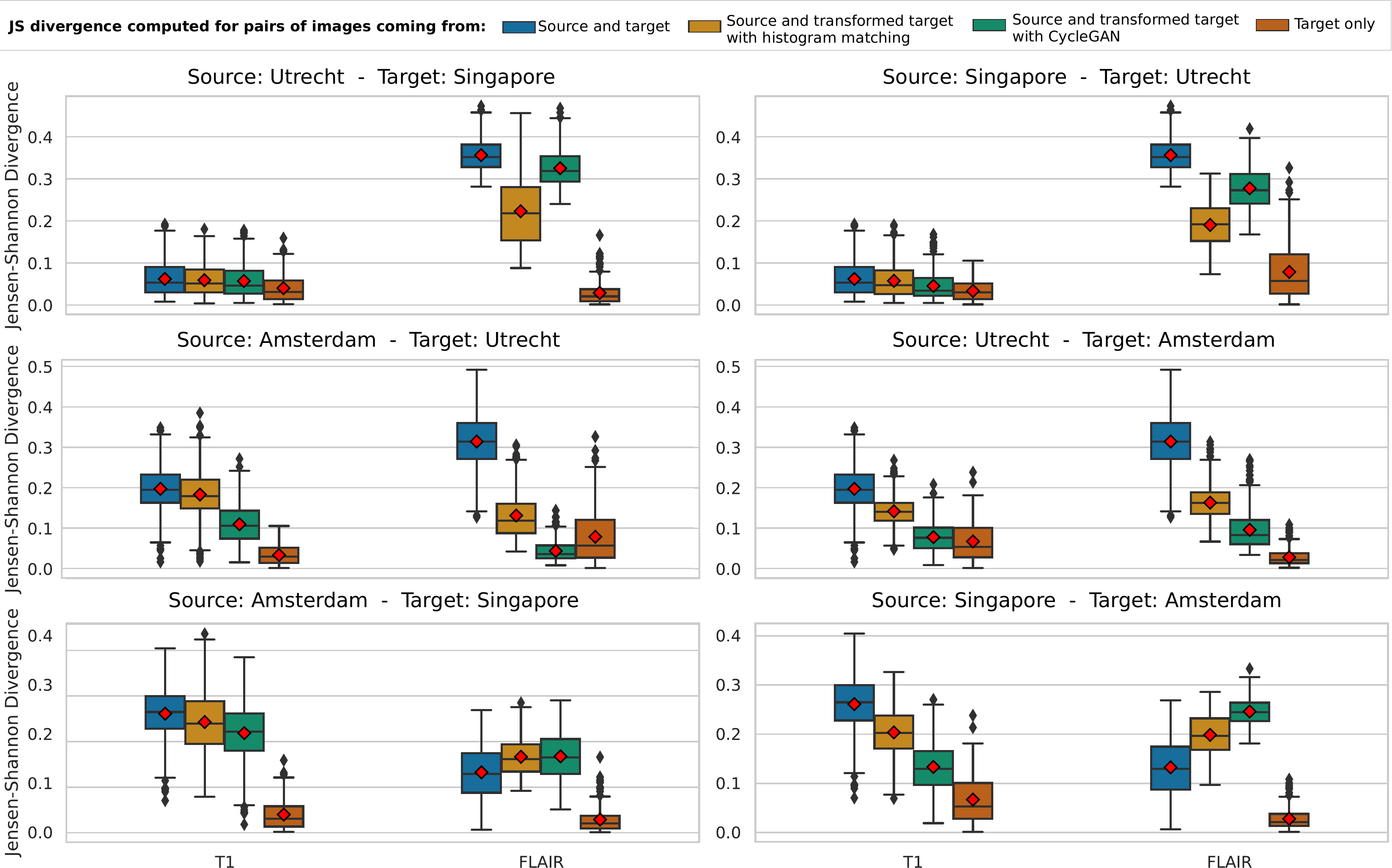}
 \caption{Co-variate shift comparison based on the JS divergence between pairs of images in 3 different scenarios: without domain adaptation (${JS}(X,Y)$, in blue), with HM domain adaptation (${JS}(X,HM(Y))$, in yellow), with CycleGAN domain adaptation (${JS}(X,F(Y))$ in green) and within the same domain (${JS}(Y,Y)$, in brown). Lower pairwise JS indicates less differences in the intensity distribution. The average divergence $\Delta_{JS}$ is shown in with a red diamond.}
 \label{fig:divergence}
\end{figure}
\noindent \textbf{Experiments and discussion.} All images were first pre-processed using z-scores normalization to account for big variations in intensity ranges. The models were implemented in Keras. We employ two different approaches to evaluate the effectiveness of CycleGAN on domain adaptation.

The first approach directly quantifies the segmentation performance with and without domain adaptation. Figure \ref{fig:dice} shows segmentation results, comparing CycleGAN and HM, but also including an upper bound given by training on images from the target domain. Figure \ref{fig:visualResults} shows some visual results from the same experiment. We use Dice coefficient \cite{dice1945measures} to measure segmentation performance. For every experiment we performed 7-fold cross validation, therefore training 7 times with 17 images (14 for training and 3 for validation) and leaving the other 3 out for testing. The results show that CycleGAN not only improves segmentation performance for all combinations of source and target but also enables segmentation in cases with null Dice before domain adaptation. When compared with HM, we observe that using CycleGAN systematically improves the mean Dice, while HM presents variable performance depending on the domain. Moreover, in all but one scenario CycleGAN outperforms HM in this task.

The second evaluation approach uses the JS pairwise divergences as a proxy to approximate the co-variate shift. Results are shown in Figure \ref{fig:divergence}, where we compare the JS divergence for multiple domains without domain adaptation (${JS}(X,Y)$), with HM domain adaptation (${JS}(X,HM(Y))$), with CycleGAN domain adaptation (${JS}(X,F(Y))$) and within the same domain (${JS}(Y,Y)$). The red diamonds in the boxplot indicate the mean pairwise JS divergence $\Delta_{JS}$ previously defined. It can be observed that, in most of the cases, CycleGAN significantly reduces $\Delta_{JS}$ outperforming the results obtained with HM.


\section{Conclusions}
In this work we show, for the first time, that CycleGAN-based domain adaptation improves lesion segmentation in multicenter brain MR images, particularly in WMH lesions. We compared the proposed approach with standard histogram matching, both in terms of segmentation quality improvement and co-variate shift between source and target domains. JS divergence is used as a measure to understand the differences between the domains, which seems to anti-correlate with segmentation performance. In other words, lower inter-domain $\Delta_{JS}$ results in better generalization from source to target. 

Our results have important practical implications. First, even if the segmentation performance is not as good as the upper bound given by training with annotated data from the target domain, it improves segmentation in cases which had null Dice before domain adaptation. This could be used to automatically detect the presence of WMH lesions in completely unseen domains without ground-truth. Second, differently from other adversarial domain adaptation techniques \cite{Kamnitsas2017,Orbes2019} which require to access the unannotated images of the target domain while training the segmentation network, our method can be used on completely unseen domains without re-training the segmenter. This makes the method useful in real clinical situations where new MR machines may arrive to a hospital once the segmentation software has been deployed.

In the future, we plan to extend our study to other type of brain lesions (e.g. stroke or brain tumours) which could also benefit from this approach.\\

\acknowledgments 
The authors gratefully acknowledge the support of UNL (CAID-PIC-50220140100084LI) and
ANPCyT (PICT 2018-03907).
 
\section{Appendix}

\subsection{CycleGAN Generator Architecture}
\label{app:generator}
For the generator, we adopted the original architecture from CycleGAN with the difference that layers are converted from 2D to 3D, extending the kernel dimensions accordingly (e.g. convolutions are $7\times7\times7$ instead of $7\times7$). Reflection padding was used to reduce artifacts in the first and last layer. The last layer does not have an activation function, since the model can adapt according to the intensity range of each domain.

\begin{table}[h!]
\centering
\begin{tabular*}{\textwidth}{c @{\extracolsep{\fill}} cccccc}

\hline
           &            & \textbf{Kernel}      & \textbf{Stride}      & \textbf{\#Kernels} & \textbf{Activation} & \textbf{Padding} \\ \hline
L1         & Conv3D     & (f:7,7,7)            & (s:1,1,1)            & (N:32)             & ReLu                & (RP: 3,3,3)      \\ \hline
L2         & Conv3D     & (f:3,3,3)            & (s:2,2,2)            & (N:64)             & ReLu                &                  \\ \hline
L3         & Conv3D     & (f:3,3,3)            & (s:2,2,2)            & (N:128)            & ReLu                &                  \\ \hline
L4 ... L13 & ResBlock   & (f:3,3,3)            & (s:1,1,1)            & (N:128)            & ReLu                &                  \\
           &            & (f:3,3,3)            & (s:1,1,1)            & (N:128)            & ReLu                &                  \\ \hline
L14        & UpSampling &                      &                      &                    &                     &                  \\
           & Conv3D     & (f:3,3,3)            & (s:1,1,1)            & (N:128)            & ReLu                &                  \\ \hline
L16        & UpSampling & \multicolumn{1}{c}{} & \multicolumn{1}{c}{} &                    &                     &                  \\
           & Conv3D     & (f:3,3,3)            & (s:1,1,1)            & (N:64)             & ReLu                &                  \\ \hline
L17        & UpSampling & \multicolumn{1}{c}{} & \multicolumn{1}{c}{} &                    &                     &                  \\
           & Conv3D     & (f:3,3,3)            & (s:1,1,1)            & (N:32)             & ReLu                &                  \\ \hline
L18        & Conv3D     & (f:7,7,)             & (s:1,1,1)            & (N:2)              & None                & (RP: 3,3,3)     

\end{tabular*}

\caption{\label{table:gen} \textbf{Detailed description of the CycleGAN generator architecture}. }
\end{table}

\subsection{CycleGAN Discriminator Architecture}
\label{app:discriminator}

The discriminator follows an architecture similar to that of PatchGAN but with 3D convolutions.

\begin{table}[h!]
\centering
\begin{tabular*}{\textwidth}{c @{\extracolsep{\fill}} cccccc}
\hline
   &        & \textbf{Kernel} & \textbf{Stride} & \textbf{\#Kernels} & \textbf{Activation} & \textbf{Normalization} \\ \hline
L1 & Conv3D & (f:4,4,4)       & (s:2,2,2)       & (N:64)             & LeakyReLu           & None                   \\ \hline
L2 & Conv3D & (f:3,3,3)       & (s:2,2,2)       & (N:128)            & LeakyReLu           & Instance               \\ \hline
L3 & Conv3D & (f:3,3,3)       & (s:2,2,2)       & (N:256)            & LeakyReLu           & Instance               \\ \hline
L4 & Conv3D & (f:3,3,3)       & (s:1,1,1)       & (N:512)            & ReLu                & Instance               \\ \hline
L5 & Conv3D & (f:4,4,4)       & (s:1,1,1)       & (N:1)              & Sigmoid             & None             
\end{tabular*}
\caption{\label{table:disc} \textbf{Detailed description of the CycleGAN discriminator architecture}.}
\end{table}

\subsection{3D U-Net Architecture}
\label{app:unet}

We adopted a modifified version of the standard U-Net architecture \cite{unet2d}. We replaced the 2D convs of the standard U-Net architecture for 3D convs. The encoding blocks consist of two convolutional layers with kernel of size $3\times3\times3$, padding = 1 and ReLU activation followed by a $2 \times 2 \times 2$ max-pooling layer. The decoding blocks have also two convolutional layers, but we use upsampling via transposed convolutions before each block. The standard U-Net uses concatenation of feature maps in the skip connections. We replaced concatenation by sum to combine the localized features of the encoding path with the input of the corresponding block from the decoding path. The last layer consists of a $1 \times 1 \times 1$ convolution with softmax to output a voxel-wise probability maps.

\begin{table}[h!]
\centering
\begin{tabular*}{\textwidth}{c @{\extracolsep{\fill}} cccccc}
\hline
    &               & \textbf{Kernel} & \textbf{Stride} & \textbf{\#Kernels} & \textbf{Activation} \\ \hline
L1  & EncodingBlock &                 &                 & (N:32)             &                     \\ \hline
L2  & EncodingBlock &                 &                 & (N:64)             &                     \\ \hline
L3  & EncodingBlock &                 &                 & (N:128)            &                     \\ \hline
L4  & EncodingBlock &                 &                 & (N:256)            &                     \\ \hline
L5  & Conv3D        & (f:3,3,3)       & (s:1,1,1)       & (N:256)            & ReLu                \\ \hline
L6  & Conv3D        & (f:3,3,3)       & (s:1,1,1)       & (N:256)            & ReLu                \\ \hline
L7  & DecodingBlock &                 &                 & (N:256)            &                     \\ \hline
L8  & DecodingBlock &                 &                 & (N:128)            &                     \\ \hline
L9  & DecodingBlock &                 &                 & (N:64)             &                     \\ \hline
L10 & DecodingBlock &                 &                 & (N:32)             &                     \\ \hline
L11 & Conv3D        & (f:1,1,1)       & (s:1,1,1)       & (N:2)              &                    
\end{tabular*}
\caption{\label{table:unet} \textbf{U-Net architecture used for WMH segmentation.}  }
\end{table}

\bibliography{report} 
\bibliographystyle{spiebib} 

\end{document}